\documentclass[preprint,12pt]{elsarticle}
\journal{Engineering fracture mechanics}
\usepackage{graphics,epsfig}
\usepackage{graphicx}
\usepackage{amssymb,amstext,amsmath}
\usepackage{booktabs}
\usepackage{natbib}
\usepackage[latin1]{inputenc}
\usepackage{epstopdf}
\begin{document}
\begin{frontmatter}

\title{Stress and dislocation distributions near a crack tip in ductile single crystals}
\author{K. C. Le$^a$\footnote{corresponding author: +49 234 32-26033, email: chau.le@rub.de.}, V. N. Tran$^b$}
\address{$^a$ Lehrstuhl f\"{u}r Mechanik - Materialtheorie, Ruhr-Universit\"{a}t Bochum, D-44780 Bochum, Germany
\\
$^b$ Department of Mechanics, Faculty of Mathematics and Computer Science, Ho Chi Minh University of Science, Viet nam}

\begin{abstract} Within the continuum dislocation theory the asymptotic analysis of the plane strain crack problem for a single crystal having only one active slip system on each half-plane is provided. The results of this asymptotic analysis show that the square root stress singularity remains valid during the plastic deformation, while the dislocation density is proportional to the stress intensity factor and distributed as the square root of the distance from the crack tip. The analytical solution for the angular distribution of the dislocation density is found. 
\end{abstract}


\end{frontmatter}

Dislocations appear to reduce energy of crystals. For crystals with cracks the high stress concentration near the crack tip causes also high energy of crystals in that region. It is therefore natural to expect that, when the load is sufficiently large, dislocations nucleate near the crack tip to reduce the stress level and by this also the energy of crystals. It is then crucial to have the correct perception of how dislocations nucleate near the crack tip. Up to now, the commonly accepted point of view is that dislocations nucleate directly at the crack tip and then glide away from it under the Peach-Koehler force \cite{Rice1974,Rice1992}. However, the analysis of crack problems reveals that the resolved shear stress is large not only at the crack tip, but also in its neighborhood. Taken this for granted, then, according to the Schmid's law, dislocations must appear simultaneously in that neighborhood exhibiting the collective character of dislocation nucleation. Since the typical dislocation density is high (about $10^8\div 10^{15}$ dislocations per square meter), it makes sense to use the continuum approach to study this problem. 

This short paper aims at finding the stress and dislocation distribution near the crack tip in ductile single crystals within the continuum dislocation theory (CDT) proposed by Berdichevsky\cite{Berdichevsky2006a,Berdichevsky2006b} and developed further in \cite{Berdichevsky-Le2007}-\cite{Le2015}. Considering the plane strain crack problem, we assume that during the plastic deformation only one slip system on each half-plane of the crystal is active. We provide an asymptotic analysis of this crack problem in the polar coordinates. The results of this asymptotic analysis show that the square root singularity for the stress field near the crack tip remains valid. This agrees with the singularity of HRR-field obtained by \citet{Hutchinson1968,Rice1968} in conventional plasticity for the materials with linear hardening. What the dislocation distribution near the crack tip is concerned, we show that they must be distributed such that the resolved shear stress is balanced with the back stress in accordance with the equilibrium of micro-forces acting on dislocations. This leads to the power law distribution $\sqrt{r}$, with $r$ being the distance from the crack tip, for the dislocation density, with its intensity being proportional to the stress intensity factor. We find also the universal angular distribution of the dislocation density. Note that the crack-tip fields in single crystal has been analyzed within the discrete dislocation dynamics in \cite{Giessen2001}. Experimental observations of the dislocation distribution near the crack tip in single crystals by electron tomography have been reported in \cite{Tanaka2008}. Another quite promising experimental method of measuring the dislocation density by using electron backscatter diffraction (EBSD) technique has been developed in \cite{Kysar2002,Kysar2010}.

\section{Plane strain crack problem for single crystal}

\begin{figure}[htb]
	\centering
	\includegraphics[width=7cm]{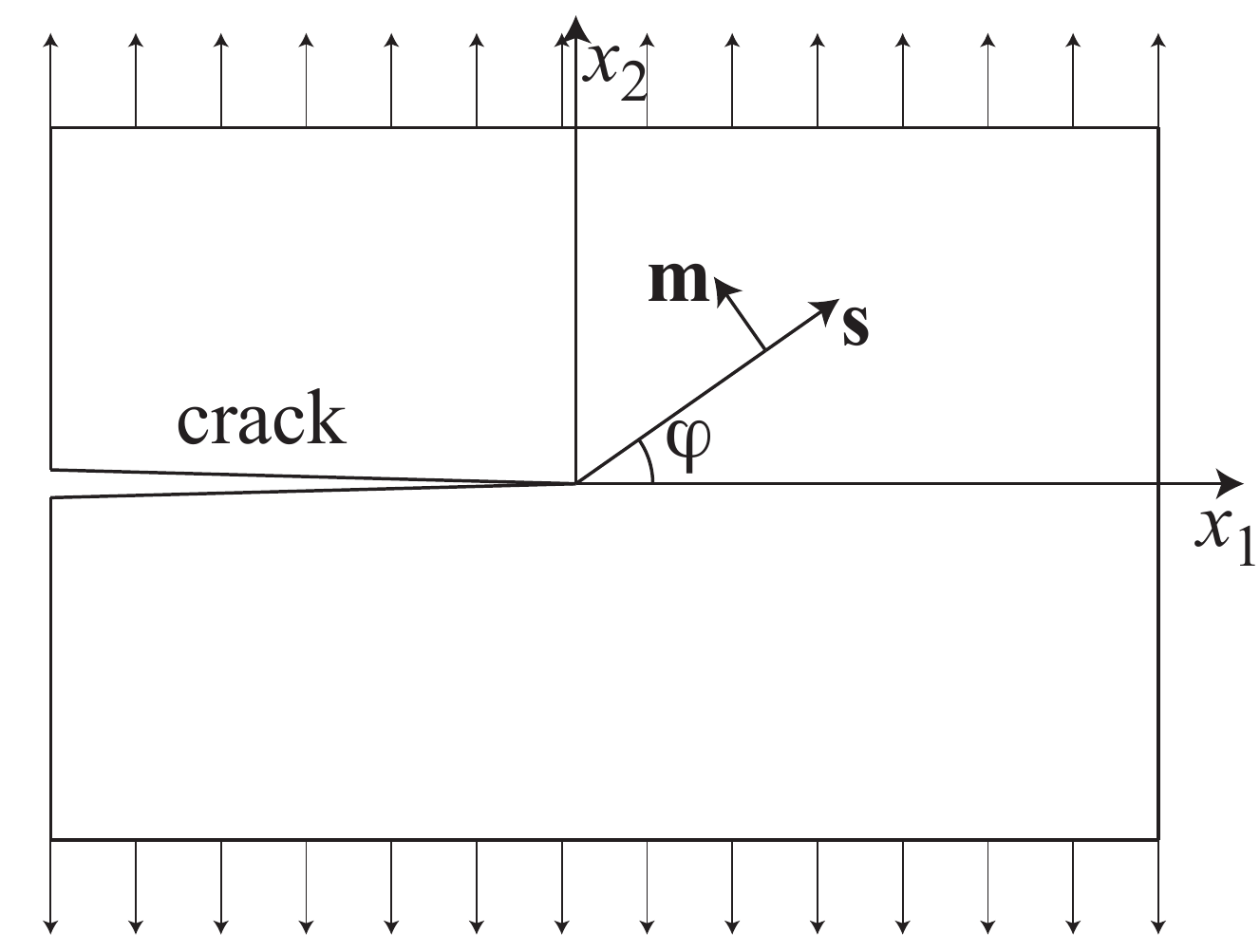}
	\caption{The plane strain crack problem for single crystal}
	\label{fig:1}
\end{figure}

Consider the plane strain problem for a single crystal containing a crack lying on the left-half of the $x_1$-axis as shown in Fig.~\ref{fig:1}. The depth of the crystal in the $x_3$-direction is taken large enough to guarantee the plane strain state having two non-zero components of displacement vector $u_1 = u_1(x_1, x_2)$ and $u_2 = u_2(x_1,x_2)$. The crystal is oriented in such a way that its lattice and mechanical properties as well as the loading condition (say tractions acting at the outer boundary) are symmetric with respect to the reflection about the $x_1$-axis. Because of this mirror symmetry it is sufficient to consider the upper-half of the crystal. If the load is small enough, then it is natural to assume that the crystal with this crack deforms elastically. However, if the load exceeds some threshold value, dislocations can occur causing the plastic deformation of the crystal. We assume that, during this plastic deformation, only one slip system from each half of the crystal is active and the dislocations are straight lines parallel to the $x_3$-axis. For the more realistic crack problems in single fcc and bcc crystals having several active slip systems the reader may consult \cite{Rice1987}. Letting $\mathbf{s}=(\cos \varphi, \sin \varphi, 0)$ denote the slip directions, and $\mathbf{m}=(-\sin \varphi, \cos \varphi, 0)$ the normal vector to the slip planes of the slip system in the upper-half of the crystal, we may express the plain strain plastic distortion tensor in the form $\boldsymbol{\beta}=\beta (x_1,x_2)\mathbf{s}\otimes \mathbf{m}$. We are going to determine the displacements $u_1(x_1,x_2)$, $u_2(x_1,x_2)$, and the plastic slip $\beta (x_1,x_2)$ near the crack tip during this plastic deformation. 

For the plane strain state the non-zero in-plane components of the symmetric strain tensor $\boldsymbol{\varepsilon }=\frac{1}{2}(\nabla \mathbf{u}+\mathbf{u}\nabla )$ are
\begin{equation*}
\varepsilon _{11}=u_{1,1},\quad \varepsilon_{12}=\varepsilon_{21}=\frac{1}{2}(u_{1,2}+u_{2,1}),\quad \varepsilon_{22}=u_{2,2}. %\label{E:totalstrain}
\end{equation*}
Throughout the paper the comma standing before an index is used to denote the partial derivative with respect to the corresponding coordinate. The in-plane components of the symmetric plastic strain tensor $\boldsymbol{\varepsilon }^p=\frac{1}{2}(\boldsymbol{\beta }+\boldsymbol{\beta }^T)$ equal
\begin{equation*}
\varepsilon^{p}_{11}=-\frac{1}{2}\beta \sin 2\varphi ,\quad \varepsilon^{p}_{12}=\varepsilon^{p}_{21}=\frac{1}{2}\beta \cos 2 \varphi ,\quad
\varepsilon^{p}_{22}=\frac{1}{2}\beta \sin 2 \varphi .
\end{equation*}
With these total and plastic strain tensors we obtain the in-plane components of the symmetric elastic strain tensor $\boldsymbol{\varepsilon }^e=\boldsymbol{\varepsilon }-\boldsymbol{\varepsilon} ^p$ in the form
\begin{equation}\label{2.1}
\begin{split}
\varepsilon^{e}_{11}=u_{1,1}+\frac{1}{2}\beta \sin 2\varphi ,\quad \varepsilon^{e}_{2,2}=u_{2,2}-\frac{1}{2}\beta \sin 2\varphi 
\\
\varepsilon^{e}_{12}=\varepsilon^{e}_{21}=\frac{1}{2}(u_{1,2}+u_{2,1}-\beta \cos 2 \varphi ) .
\end{split}
\end{equation}
Let us compute Nye-Bilby-Kr\"oner's dislocation density tensor (introduced in \cite{Nye1953}-\cite{Kroener1955}) $\boldsymbol{\alpha }=-\boldsymbol{\beta }\times \nabla $, with $\times$ being the vector product. For the plane strain plastic slip $\beta (x_1,x_2)$ there are two non-zero components of this tensor given by
\begin{equation*}
\alpha _{13}  =  (\beta _{,1}\cos \varphi +\beta _{,2}\sin \varphi )\cos \varphi ,\quad
\alpha _{23}  =  (\beta _{,1}\cos \varphi +\beta _{,2}\sin \varphi )\sin \varphi .
\end{equation*}
These are the component of the net Burgers' vector of all excess dislocations whose dislocation lines cut the area perpendicular to the $x_3$-axis. Thus, the net Burgers' vector of excess dislocations shows in the slip direction $\mathbf{s}$ indicating that we are dealing with the edge dislocations only. The scalar dislocation density (or the number of dislocations per unit area) equals
\begin{equation}\label{2.2}
\rho =\frac{1}{b}\sqrt{\alpha ^2_{13}+\alpha ^2_{23}}=\frac{1}{b}|\beta _{,1}\cos \varphi +\beta _{,2}\sin \varphi |=\frac{1}{b}|\partial _s \beta |,
\end{equation}
where $\partial _s \beta =\nabla \beta \cdot \mathbf{s}$. Similar quantities in the lower-half plane can be obtained if we replace in the above formulas $\varphi $ by $-\varphi $.

For most metals the elastic strain tensor $\boldsymbol{\varepsilon }^e$ is usually small. Therefore the free energy density per unit volume of the crystal with continuously distributed dislocations can be proposed in the form \citep{Berdichevsky2006a,Berdichevsky2006b}
\begin{equation}\label{2.3}
\psi (\boldsymbol{\varepsilon }^e,\rho )=\frac{1}{2}\lambda (\text{tr}\boldsymbol{\varepsilon }^e)^2+\mu \boldsymbol{\varepsilon }^e\mathbf{:}\boldsymbol{\varepsilon }^e +\mu k \ln \frac{1}{ 1-\frac{\rho }{\rho_{s}}},
\end{equation}
with $\lambda $ and $\mu $ the Lam\'{e}'s constants, $\rho _s$ the saturated dislocation density, and $k$ the material constant. The first two terms in \eqref{2.3} represents the contribution to the energy due to the elastic strain, while the last term corresponds to the energy of the dislocation network. The logarithmic energy term stems from two facts: (i) energy of the dislocation network for small dislocation densities is the sum of energy of non-interacting dislocations (see the reasonings based on the statistical mechanics of dislocations in \citep{Le2001,Le2011}, and (ii) there exists a saturated dislocation density which characterizes the closest packing of dislocations admissible in the discrete crystal lattice. The logarithmic term \citep{Berdichevsky2006b} ensures a linear increase of the energy for small dislocation density $\rho$ and tends to infinity as $\rho$ approaches the saturated dislocation density $\rho_s$ hence providing an energetic barrier against over-saturation. For small up to moderate dislocation densities this logarithmic term can be replaced by its Taylor expansion
\begin{equation*}
%\label{2.3a}
\ln \frac{1}{ 1-\frac{\rho }{\rho_{s}}}\simeq \frac{\rho }{\rho_{s}}+\frac{1}{2}\frac{\rho ^2}{\rho _s^2}.
\end{equation*}
For simplicity of the subsequent analysis we shall use this approximation only. Taking all the above formulas into account, we write the bulk energy density per unit volume of the crystal with continuously distributed dislocations in the form
\begin{multline*}%\label{2.4}
\psi (\boldsymbol{\varepsilon }^e,\rho )=\frac{1}{2}\lambda (u_{1,1}+u_{2,2})^2+\mu (u_{1,1}+\frac{1}{2}\beta \sin 2\varphi )^2+\mu (u_{2,2}-\frac{1}{2}\beta \sin 2\varphi )^2
\\
+\frac{1}{2}\mu (u_{1,2}+u_{2,1}-\beta \cos 2\varphi )^2+\mu k (\frac{|\partial _s \beta |}{b\rho_{s}}+\frac{1}{2}\frac{(\partial _s \beta )^2}{b^2\rho _s^2}).
\end{multline*}
The energy functional per unit depth of the crystal becomes
\begin{multline}\label{2.5}
I[u_\alpha ,\beta ]=\int_{\Omega } \biggl[ \frac{1}{2}\lambda (u_{1,1}+u_{2,2})^2+\mu (u_{1,1}+\frac{1}{2}\beta \sin 2\varphi )^2+\mu (u_{2,2}-\frac{1}{2}\beta \sin 2\varphi )^2
\\
+\frac{1}{2}\mu (u_{1,2}+u_{2,1}-\beta \cos 2\varphi )^2+\mu k (\frac{|\partial _s \beta |}{b\rho_{s}}+\frac{1}{2}\frac{(\partial _s \beta )^2}{b^2\rho _s^2}) \,\biggr] \, dx, 
\end{multline}
where $\Omega $ is the domain occupied by the undeformed crystal having the crack and $dx=dx_1dx_2$. Since we are interested in the fields near the crack tip, the virtual work of the traction acting at the outer boundary is dropped in this functional. We require that the displacements $u_1$ and $u_2$ and the plastic slip $\beta $ be continuous on the positive $x_1$-axis. On the contrary, on the crack faces $x_1<0$, $x_2=\pm 0$ (which are assumed to be not in contact with each other in the deformed state for the mode I crack) no constraints are imposed on $u_1$, $u_2$, and $\beta $ (free boundary). Provided the dissipation caused by the dislocation motion is negligible, then the true displacements $u_1$, $u_2$ and plastic slip $\beta$ in the {\it final} state of equilibrium minimize energy functional \eqref{2.5} among all admissible displacements and plastic slips satisfying the above boundary conditions.

Let us derive the equilibrium equations and boundary conditions for the true displacement vector and the plastic slip. For plane strain problems it is convenient to use the index notation with the greek indices running from 1 to 2 and with the summation convention being applied for two-dimensional components of vectors and tensors. Then, for instance, the elastic strain tensor has the following non-zero components
\begin{equation*}
\varepsilon ^e_{\alpha \beta }=\frac{1}{2}(u_{\alpha ,\beta }+u_{\beta ,\alpha })-\frac{1}{2}\beta (s_\alpha m_\beta +s_\beta m_\alpha ).
\end{equation*}
Keeping this index notation in mind, we compute now the first variation of the energy functional \eqref{2.5}
\begin{equation*}
\delta I=\int_{\Omega }[\sigma _{\alpha \beta }(\delta u_{\alpha ,\beta }-\delta \beta s_\alpha m_\beta )+\kappa _\alpha \delta \beta _{,\alpha }]\, dx,
\end{equation*}
where
\begin{equation}\label{3.1}
\begin{split}
\sigma _{\alpha \beta }=\frac{\partial \psi }{\partial \varepsilon ^e_{\alpha \beta }}=\lambda \varepsilon ^e_{\gamma \gamma }\delta _{\alpha \beta }+2\mu \varepsilon ^e_{\alpha \beta },
\\
\kappa _\alpha =\frac{\partial \psi }{\partial \beta_{,\alpha }}=\mu k\left[ \frac{1}{b\rho _s}\text{sign}(\partial _s\beta )+\frac{\partial _s\beta }{b^2\rho _s^2}\right] s_\alpha .
\end{split}
\end{equation}
We call $\sigma _{\alpha \beta }$ the (symmetric) Cauchy stress tensor, and $\kappa _\alpha $ the higher order stress vector. Integrating by parts and omitting the surface integral at the outer boundary, we obtain
\begin{equation*}
\delta I=\int_{\Omega }[-\sigma _{\alpha \beta ,\beta }\delta u_\alpha -(\tau +\kappa _{\alpha ,\alpha })\delta \beta ]\, dx-\int_{x_1<0}(\sigma _{\alpha 2}\delta u_\alpha +\kappa _2\delta \beta)|^{x_2=+0}_{x_2=-0}dx_1,
\end{equation*}
where $\tau =\sigma _{\alpha \beta }s_\alpha m_\beta $ is the resolved shear stress (or Schmid stress). Thus, the equation $\delta I=0$, together with the arbitrariness of $\delta u_\alpha $ and $\delta \beta $, implies the equilibrium equations of macro-forces acting on the material volume element
\begin{equation}
\label{3.2}
\sigma _{\alpha \beta ,\beta }=0, 
\end{equation}
and of micro-forces acting on dislocations
\begin{equation}
\label{3.2a}
\tau +\kappa _{\alpha ,\alpha }=0
\end{equation}
in $\Omega $, as well as
\begin{equation}
\label{3.3}
\sigma _{\alpha 2}=0, \quad \kappa _2=0 \quad \text{for $x_1<0$, $x_2=\pm0$}.
\end{equation}
In addition to these we can pose the boundary condition
\begin{equation}
\label{3.3a}
\beta (x_1,0)=0 \quad \text{for $x_1>0$}
\end{equation}
which is simply the consequence of the mirror symmetry of the crack problem. 
The system of equations \eqref{3.1}, \eqref{3.2}, \eqref{3.2a} together with the boundary conditions \eqref{3.3} and \eqref{3.3a} constitute the local crack problem to determine the displacements $u_\alpha $ and the plastic slip $\beta $. 

To get the governing equations in terms of displacements and plastic slip we substitute the constitutive equations \eqref{3.1}, with $\varepsilon ^e_{\alpha \beta }$ and $\rho $ from \eqref{2.1} and \eqref{2.2}, into the equilibrium equations \eqref{3.2} and \eqref{3.2a}. This leads to
\begin{equation}\label{3.4}
(\lambda+\mu)u_{\beta ,\beta \alpha }+\mu u_{\alpha ,\beta \beta }-\mu\beta_{,\beta }(s_\alpha m_\beta +s_\beta m_\alpha )=0, 
\end{equation}
and 
\begin{equation}\label{3.5}
\mu (u_{\alpha ,\beta }+u_{\beta ,\alpha })s_\alpha m_\beta -\mu \beta+\frac{\mu k}{b^2\rho_s^2}\beta_{,\alpha \beta }s_\alpha s_\beta =0. 
\end{equation}

\section{Asymptotic analysis}
Near the crack tip different terms in the governing equations \eqref{3.4} and \eqref{3.5} will have different orders of smallness. It is convenient to do the asymptotic analysis of these equations in the polar coordinate system $r$ and $\theta $, where
\begin{equation*}
	x_1=r\cos\theta ,\quad
	x_2=r\sin\theta .
\end{equation*}
Let us assume the asymptotically main terms for displacements and plastic slip near the crack tip in the form
\begin{equation}
\label{3.6}
u_\alpha (r,\theta )=r^n f_\alpha (\theta ),\quad \beta (r,\theta )=r^m g(\theta ),
\end{equation}
where $f_\alpha (\theta )$ and $g(\theta )$ are unknown functions describing the angular distribution of the displacements and plastic slip, and $n$ and $m$ are unknown numbers. Since the displacements and the plastic slip cannot be singular at the crack tip, both $m$ and $n$ must be positive. Besides, based on the solutions of other similar crack problems we may assume that $n$ lies in the interval $(0,1)$. We substitute asymptotic formulas \eqref{3.6} into the governing equations and use the transformation rules 
\begin{equation*}
	\left[\begin{array}{l}
	\partial_1\\
	\partial_2
	\end{array}\right]=\left[\begin{array}{cr}
	\cos\theta\quad&-\frac{1}{r}\sin\theta\\
	\sin\theta\quad&\frac{1}{r}\cos\theta
	\end{array}\right]\left[\begin{array}{l}
	\partial_r\\
	\partial_\theta
	\end{array}\right]
\end{equation*}
to compute the derivatives. According to these transformation rules, the first derivatives reduce the power of $r$ by one, and the second derivatives by two. Thus, the asymptotically main terms in \eqref{3.4} are the first two because the last one has the order $r^{m-1}$ which is small compared with $r^{n-2}$. Neglecting this last term in equations \eqref{3.4} we get Lam\'{e}'s equations of linear elasticity that are uncoupled from equation \eqref{3.5}. Likewise, the boundary conditions \eqref{3.3}$_1$, after neglecting the last small term containing $\beta $, reduces to
\begin{equation*}
\lambda u_{\gamma ,\gamma }\delta _{\alpha 2}+\mu (u_{\alpha ,2}+u_{2,\alpha })=0,
\end{equation*}
so we get the plane strain elastic crack problem that can be solved by using the Airy stress function. The well-known solution of this problem shows that $n=1/2$ (see, e.g., \cite{Le2011}). Besides, for mode I crack
\begin{equation*}
\begin{split}
u_1=\frac{K_I}{\mu }\sqrt{\frac{r}{2\pi }}\cos \frac{\theta
}{2}\left(1-2\nu +\sin ^2\frac{\theta }{2}\right),
\\
u_2=\frac{K_I}{\mu }\sqrt{\frac{r}{2\pi }}\sin \frac{\theta
}{2}\left(2-2\nu -\cos ^2\frac{\theta }{2}\right),
\end{split}
\end{equation*}
and
\begin{equation}\label{3.6a}
\begin{split}
\sigma _{11}=\frac{K_I}{\sqrt{2\pi r}}\cos \frac{\theta }{2}
\left[1-\sin \frac{\theta }{2} \sin \frac{3\theta }{2} \right],\\
\sigma _{22}=\frac{K_I}{\sqrt{2\pi r}}\cos \frac{\theta }{2}
\left[1+\sin \frac{\theta }{2} \sin \frac{3\theta }{2} \right],\\
\sigma _{12}=\frac{K_I}{\sqrt{2\pi r}}\cos \frac{\theta }{2}
\sin \frac{\theta }{2} \cos \frac{3\theta }{2} .
\end{split}
\end{equation}
Here $K_I$ is the stress intensity factor that can be found only after solving the global crack problem. Since crystals with continuously distributed dislocations exhibit a linear work hardening as shown in \cite{Berdichevsky-Le2007}-\cite{Kochmann09a}, this square root stress singularity agrees with the classical result obtained by \citet{Hutchinson1968,Rice1968} in conventional plasticity. Note also that the account of dissipation as proposed in \cite{Berdichevsky-Le2007}-\cite{Kochmann09a} does not affect this result.

We turn now to equation \eqref{3.5}. If $m\in (0,1]$, then the asymptotically principal term in this equation is obviously the last one, because it has the order $r^{m-2}$ which is much larger than the orders of the first two. Neglecting the small terms in \eqref{3.5} we get a simple equation
\begin{equation*}
%\label{3.7}
\frac{\mu k}{b^2\rho_s^2}\frac{\partial ^2}{\partial s^2}\beta =0.
\end{equation*}
The continuous solution of this equation
\begin{equation*}%\label{3.8}
\beta =b\rho _0s,
\end{equation*}
with $s$ being the distance from the $x_1$-axis in the $\mathbf{s}$-direction and $\rho _0$ a constant dislocation density, would satisfy the equation and the boundary condition \eqref{3.3a} identically, but violate the boundary condition
\begin{equation*}
\kappa _2=\mu k\left[ \frac{1}{b\rho _s}\text{sign}(\partial _s\beta )+\frac{\partial _s\beta }{b^2\rho _s^2}\right] s_2=0
\end{equation*}
on the crack faces. This is not surprising, because we know that the free boundary attracts dislocations, so the constant dislocation distribution cannot stay in equilibrium near the crack faces. Therefore $m$ cannot lie in the segment $(0,1]$. For $m\in (1,2)$ the back-stress must be balanced with the resolved shear stress (having the power $r^{-1/2}$) to guarantee the equilibrium of micro-forces acting on dislocations. Therefore $m=3/2$ and $\beta =r^{3/2}g(\theta )$. Now we substitute this Ansatz into equation \eqref{3.5} and maintain the terms of order $r^{-1/2}$ in it. Then equation \eqref{3.5} becomes
\begin{equation}\label{3.9}
\sigma _{\alpha \beta }s_\alpha m_\beta + \frac{\mu k}{b^2 \rho_s^2}\left(\beta_{,11}s_1^2+2\beta_{,12}s_1s_2+\beta_{,22}s_2^2\right)=0,
\end{equation}
where the stress components must be taken from \eqref{3.6a}. To compute the back stress we use the following formulas
\begin{equation*}
\begin{split}
	\beta_{,11}&=r^{-1/2}\left\{\frac{3}{4}\left[\frac{3}{2}-\frac{1}{2}\cos2\theta\right]g(\theta)-\frac{1}{2}\sin2\theta g^{\prime}(\theta)+\sin^{2}(\theta)g^{\prime\prime}(\theta)\right\} ,\\
	\beta_{,12}&=r^{-1/2}\left[-\frac{3}{8}\sin2\theta g(\theta)+\frac{1}{2}\cos2\theta g^{\prime}(\theta)-\frac{1}{2}\sin2\theta g^{\prime\prime}(\theta)\right] ,\\
	\beta_{,22}&=r^{-1/2}\left\{\frac{3}{4}\left[\frac{3}{2}+\frac{1}{2}\cos2\theta\right]g(\theta)+\frac{1}{2}\sin2\theta g^{\prime}(\theta)+\cos^2(\theta)g^{\prime\prime}(\theta)\right\} .
\end{split}
\end{equation*}
Substituting the resolved shear stress and back stress into \eqref{3.9} and simplifying this equation we obtain 
\begin{multline*}
\frac{b^2\rho_s^2 K_I }{\mu k\sqrt{2\pi }}\cos (3\theta /2-2\varphi )\sin \theta 
+\frac{3}{2}\left[\frac{3}{2}-\frac{1}{2}\cos(2\theta-2\varphi)\right]g(\theta) \\
 -\sin(2\theta-2\varphi)g^{\prime}(\theta)+2\sin^2(\theta-\varphi)g^{\prime\prime}(\theta)=0.
\end{multline*}
Since this equation is linear, its solution $g(\theta )$ must be proportional to the factor $b^2\rho _s^2K_I/(\mu k\sqrt{2\pi })$. Therefore the problem reduces to solving the equation
\begin{multline}\label{3.11}
\cos (3\theta /2-2\varphi )\sin \theta 
+\frac{3}{2}\left[\frac{3}{2}-\frac{1}{2}\cos(2\theta-2\varphi)\right]h(\theta) \\
 -\sin(2\theta-2\varphi)h^{\prime}(\theta)+2\sin^2(\theta-\varphi)h^{\prime\prime}(\theta)=0.
\end{multline}
The inhomogeneous equation \eqref{3.11} must be subjected to the boundary conditions
\begin{equation}
\label{3.12}
h(0)=0,\quad \frac{3}{2}\cos \varphi \, h(\pi )+\sin \varphi \, h^\prime (\pi )=0.
\end{equation}
The second boundary condition means nothing else but the vanishing dislocation density near the crack faces. Only in this case the equilibrium of dislocations is guaranteed. 

To solve equation \eqref{3.11} we make the change of unknown function
\begin{equation*}
h(\theta )=\sqrt{\sin(\theta -\varphi )} \, y(\theta ) \quad \text{for $\theta >\varphi $}.
\end{equation*}
It is straightforward to show that equation \eqref{3.11}, in terms of $y(\theta )$, becomes
\begin{equation}
\label{3.14}
y^{\prime \prime }+y=-\frac{\cos(3\theta /2-2\varphi )\sin \theta }{2(\sin(\theta -\varphi ))^{5/2}}.
\end{equation}
The general solution of \eqref{3.14} that can directly be verified, reads
\begin{equation*}
y(\theta )=-\frac{1}{3\sqrt{\sin(\theta -\varphi )}}[-3\sin \frac{\theta }{2}+\sin \frac{3\theta }{2}-4\sin (\frac{3\theta }{2}-2\varphi )]+C_1\cos \theta +C_2\sin \theta .
\end{equation*}
For $\theta <\varphi $ the change of unknown function $h(\theta )=-\sqrt{\sin(\varphi -\theta )} \, y(\theta )$ would do the job in that interval. Then we must combine the solutions in two intervals taking into account the continuity of the dislocation density and the analytic continuation of $\beta $ across the singular point $\theta =\varphi$. Returning to the original function $h(\theta )$ we obtain
\begin{equation*}
h(\theta )=\begin{cases}
  p(\theta )+(C_1\cos \theta +C_2\sin \theta )\sqrt{\sin(\theta -\varphi )}    & \text{for $\theta >\varphi $}, \\
  p(\theta )-(C_1\cos \theta +C_2\sin \theta )\sqrt{\sin(\varphi -\theta )}    & \text{for $\theta <\varphi $},
\end{cases}
\end{equation*}
where 
\begin{equation*}
p(\theta )=-\frac{1}{3}[-3\sin \frac{\theta }{2}+\sin \frac{3\theta }{2}-4\sin (\frac{3\theta }{2}-2\varphi )] .
\end{equation*}

\begin{figure}[htb]
	\centering
	\includegraphics[width=8cm]{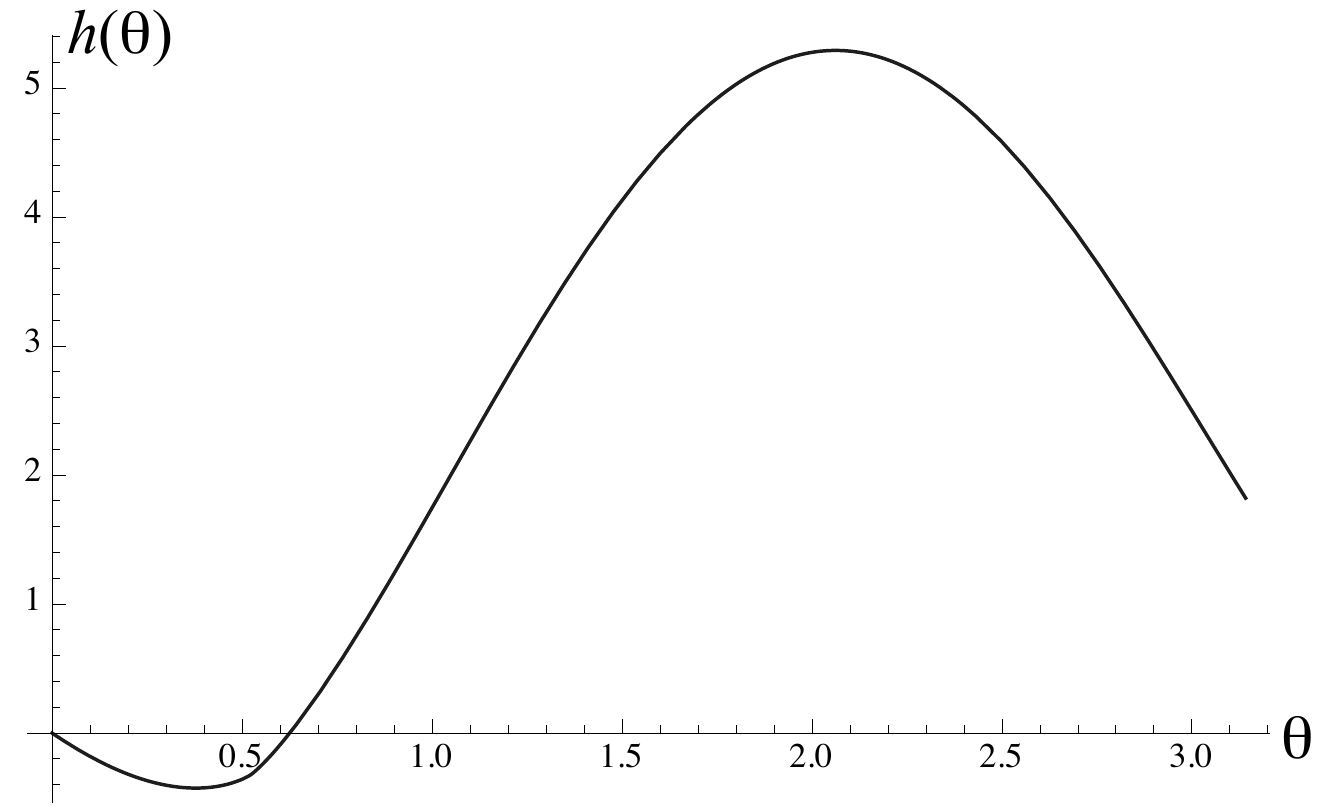}
	\caption{Angular distribution of the plastic slip}
	\label{fig:2}
\end{figure}

The unknown coefficients $C_1$ and $C_2$ can be found from the boundary conditions \eqref{3.12} yielding 
\begin{equation*}
C_1=-\frac{4\sin 2\varphi }{3\sqrt{\sin \varphi }}, \quad C_2=\frac{4\sin 2\varphi }{3\sqrt{\sin \varphi }}\cot \varphi .
\end{equation*}
Thus, the final solution reads
\begin{equation}
\label{3.15}
h(\theta )=\begin{cases}
  p(\theta )-\frac{4\sin 2\varphi }{3\sqrt{\sin \varphi }}(\cos \theta -\cot \varphi \, \sin \theta )\sqrt{\sin(\theta -\varphi )}    & \text{for $\theta >\varphi $}, \\
  p(\theta )+\frac{4\sin 2\varphi }{3\sqrt{\sin \varphi }}(\cos \theta -\cot \varphi \, \sin \theta )\sqrt{\sin(\varphi -\theta )}    & \text{for $\theta <\varphi $}.
\end{cases}
\end{equation}
The plot of function $h(\theta )$ (for $\varphi =\pi /6$) is shown in Fig.~\ref{fig:2}.

\begin{figure}[htb]
	\centering
	\includegraphics[width=8cm]{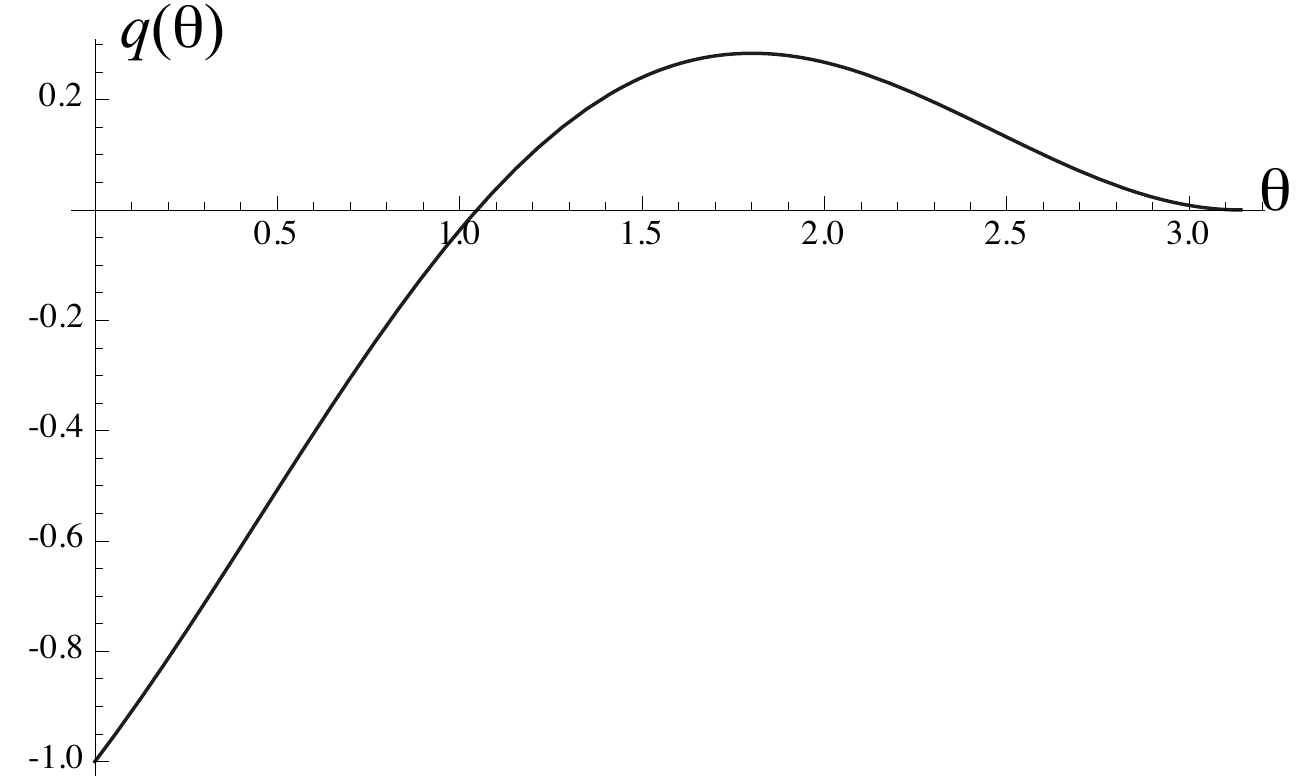}
	\caption{Angular distribution of the dislocation density}
	\label{fig:3}
\end{figure}

To summarize, the asymptotic formula for $\beta (r,\theta )$ is
\begin{equation*}
\beta (r,\theta )=\frac{b^2\rho _s^2K_I}{\mu k\sqrt{2\pi }}r^{3/2}h(\theta ),
\end{equation*}
with $h(\theta )$ from \eqref{3.15}. Differentiating this formula, we find the signed dislocation density in the simple form
\begin{equation}
\label{3.16}
\rho =\frac{\partial _s\beta }{b} =\frac{b\rho _s^2K_I}{\mu k\sqrt{2\pi }}r^{1/2} 2\cos ^2(\theta /2)\sin ((\theta -2\varphi )/2) .
\end{equation}
We see that the dislocation density is proportional to the stress intensity factor and distributed as $\sqrt{r}$ in the radial direction. Fig.~\ref{fig:3} (again for $\varphi =\pi /6$) shows the angular distribution $q(\theta )=2\cos ^2(\theta /2)\sin ((\theta -2\varphi )/2)$ of the dislocation density.

\section{Conclusion}
We have shown in this paper that, within the continuum dislocation theory, the crack in ductile single crystals causes the square root stress singularity even during the plastic deformation. The dislocation density is proportional to the stress intensity factor and distributed in accordance with formula \eqref{3.16}. The near-crack-tip fields for the displacements and plastic slip can be used to design singular crack-tip elements to solve crack problems in ductile single crystals by the finite element method. It remains still unclear, whether some threshold in loading exists for the onset of dislocations nucleation. To answer this question one needs to compare the energy of the crystal containing a crack without and with dislocations. Another open issue is the determination of the near-crack tip fields for single crystals with several active slip systems. These issues, as well as the question of the crack growth, will be addressed in our forthcoming papers. Last but not least, it would be quite convincing if this theoretical result for the dislocation distribution could be compared with the experimental observations and measurements. It is hoped that this paper would serve as motivation for experimentalists using EBSD-technique to measure the dislocation density near the crack tip.

\bigskip
\noindent {\it Acknowledgments}

The financial support by the German Science Foundation (DFG) through the research projects LE 1216/4-2 and GP01-G within the Collaborative Research Center 692 (SFB692) is gratefully acknowledged.

\end{document}